\documentclass[twocolumn]{aastex62}

\graphicspath{{./}{figures/}}

\received{XXX, 2018}
\revised{XXX, 2019}
\accepted{XXX, 2019}
\submitjournal{ApJ}

\shorttitle{QPP in X8.2 Flare}
\shortauthors{Hayes et al.}
\begin{document}

\title{Persistent Quasi-Periodic Pulsations During a Large X-Class Solar Flare}

%\correspondingauthor{Laura A. Hayes}
%\email{hayesla@tcd.ie}

\author{Laura A. Hayes}
\affil{School of Physics, Trinity College Dublin, Dublin 2, Ireland}
\affiliation{Solar Physics Laboratory, Code 671, Heliophysics Science Division, NASA Goddard Space Flight Center, Greenbelt, MD 20771, USA}
\affiliation{Adnet Systems, Inc}

\author{Peter T. Gallagher}
\affil{School of Physics, Trinity College Dublin, Dublin 2, Ireland}
\affil{School  of  Cosmic  Physics,  Dublin  Institute  for  Advanced Studies,  Dublin,  D02XF85,  Ireland.}

\author{Brian R. Dennis}
\affiliation{Solar Physics Laboratory, Code 671, Heliophysics Science Division, NASA Goddard Space Flight Center, Greenbelt, MD 20771, USA}

\author{Jack Ireland}
\affiliation{Adnet Systems, Inc}
\affiliation{Solar Physics Laboratory, Code 671, Heliophysics Science Division, NASA Goddard Space Flight Center, Greenbelt, MD 20771, USA}

\author{Andrew Inglis}
\affiliation{Solar Physics Laboratory, Code 671, Heliophysics Science Division, NASA Goddard Space Flight Center, Greenbelt, MD 20771, USA}

\author{Diana E. Morosan}
\affiliation{Department of Physics, University of Helsinki, P.0. Box 64, Helsinki, Finland.}

%% Mark off the abstract in the ``abstract'' environment. 
\begin{abstract}
Solar flares often display pulsating and oscillatory signatures in the emission, known as quasi-periodic pulsations (QPP). QPP are typically identified during the impulsive phase of flares, yet in some cases, their presence is detected late into the decay phase. Here, we report extensive fine structure QPP that are detected throughout the large X8.2 flare from 2017 September 10. Following the analysis of the thermal pulsations observed in the GOES/XRS and the 131 \AA\ channel of SDO/AIA, we find a pulsation period of $\sim$65~s during the impulsive phase followed by lower amplitude QPP with a period of $\sim$150~s in the decay phase, up to three hours after the peak of the flare. We find that during the time of the impulsive QPP, the soft X-ray source observed with RHESSI rapidly rises at a velocity of approximately 17~kms$^{-1}$ following the plasmoid/coronal mass ejection (CME) eruption. We interpret these QPP in terms of a manifestation of the reconnection dynamics in the eruptive event. During the long-duration decay phase lasting several hours, extended downward contractions of collapsing loops/plasmoids  that reach the top of the flare arcade are observed in EUV. We note that the existence of persistent QPP into the decay phase of this flare are most likely related to these features. The QPP during this phase are discussed in terms of MHD wave modes triggered in the post-flaring loops.

\end{abstract}

%% Keywords should appear after the \end{abstract} command. 
%% See the online documentation for the full list of available subject
%% keywords and the rules for their use.

%% From the front matter, we move on to the body of the paper.
%% Sections are demarcated by \section and \subsection, respectively.
%% Observe the use of the LaTeX \label
%% command after the \subsection to give a symbolic KEY to the
%% subsection for cross-referencing in a \ref command.
%% You can use LaTeX's \ref and \label commands to keep track of
%% cross-references to sections, equations, tables, and figures.
%% That way, if you change the order of any elements, LaTeX will
%% automatically renumber them.
%%
%% We recommend that authors also use the natbib \citep
%% and \citet commands to identify citations.  The citations are
%% tied to the reference list via symbolic KEYs. The KEY corresponds
%% to the KEY in the \bibitem in the reference list below. 

\section{Introduction} \label{sec:intro}
A common characteristic in solar and stellar flaring emission is the presence of oscillatory or pulsating signatures known as quasi-periodic pulsations (QPP). QPP have typical periods ranging from seconds to several minutes, and their presence in both a solar and stellar context has been widely discussed in the literature (see \cite{nak09} and \cite{vandoor2016} for recent comprehensive reviews). 

QPP are prominently identified during the impulsive phase of flares in non-thermal hard X-ray and microwave observations \citep{parksandwink,  fleishman, inglisdennis, huang, hayes2016}.  However, recent studies have provided evidence of QPP signatures across the whole electromagnetic spectrum of flaring emissions (e.g.  $\gamma$-rays \citep{nak10}, Ly$\alpha$ \citep{milligan17}, chromospheric line emission \citep{brosius} and doppler shift velocities \citep{tian},  soft X-ray and EUV \citep{dolla2012, simoes, dennis2017, dominique2018} and decimetric radio bursts \citep{li, kupriyanova16}). 
Given that QPP are directly linked to all aspects of flaring energy release, their observations provide clues to the physical processes operating in the flare site. 

Despite the increased number of observations, the nature and underlying physical mechanism for the generation of QPP remain debated. Various mechanisms have been suggested regarding possible causes of QPP in solar and stellar flares (see \cite{mclaughlin} for a recent review of QPP models). These mechanisms can be categorized into either \textit{oscillatory} or \textit{self-oscillatory} processes. 

In the oscillatory category, QPP are described in terms of motions around an equilibrium~--- such as magnetohydrodynamic (MHD) oscillations in or near the energy release site. The observed periodicities of QPP are consistent with the expected timescales of MHD waves in coronal conditions, making them an attractive explanation. For example, the decayless regime of kink mode oscillations \citep[e.g.][]{decayless_kink1, decayless_kink2}, now identified as a common phenomenon in flaring active region, have timescales that overlap with that of flaring QPP and may explain the presence of low-amplitude QPP that exist for several cycles of oscillation. MHD waves propagating in flaring loops could periodically modulate the emission directly or affect charged particle dynamics \citep{nak09}. MHD waves can also play a role in the periodic triggering of magnetic reconnection \citep{nak06, nak11}. The observed QPP period would depend on the properties of the flaring loops and the type of MHD wave mode (e.g. sausage, kink, slow magneto-acoustic).  When interpreted in terms of MHD wave modes, QPP observations offer an opportunity to perform coronal seismology - the inference of magnetic and plasma properties from observed oscillatory behavior in the corona \citep{nak09, demoortel}.

In the self-oscillatory case, the QPP are connected to the dynamics of magnetic reconnection and energy release, which may be time-dependent. Numerical MHD simulations have shown that reconnection does not proceed in a steady fashion, but instead exhibits a time-dependent behavior, producing periodic outputs from an aperiodic driver \citep{murray09, mclaughlin2012, thurgood17}. Similarly,  the magnetic island reconnection system \citep{drake, guidoni} induces a repetitive regime of reconnection resulting in a periodic acceleration of electrons into flaring loops. The QPP signatures in flaring lightcurves are then a manifestation of periodic energy release, and their period is related to the timescale of oscillatory reconnection or the generation and interaction of magnetic islands in current sheets. 

To date, observations have not allowed for a definitive choice between possible mechanisms. It is quite likely that different processes operate in different cases and during different phases of the flare. Recent work has shown that thermal soft X-ray QPP are a prominent feature in flaring emissions, clearly identified during the impulsive phase in the detrended or time-derivative lightcurves \citep{dolla2012, simoes, hayes2016, kolotkov18}. In some cases the pulsations are observed to persist late into the decay phase \citep{dennis2017, hayes2016}. These fine structure QPP must be linked to some aspects of the flaring region and their relationship to other flaring parameters requires investigation. 

In this paper we investigate the nature of extensive QPP detected in the thermal emissions from the well-observed X8.2 solar flare on September 10, 2017. By using a combination of high cadence soft X-ray measurements, together with spatially resolved observations using the Atmospheric Imaging Assembly (AIA) onboard the \textit{Solar Dynamics Observatory (SDO)} \citep{lemen} and the \textit{Reuven Ramaty High Energy Solar Spectroscopic Imager (RHESSI)} \citep{lin}, we relate the characteristics of the detected QPP to the length scales of the flaring event.

\begin{figure*}
\centering
\includegraphics[width = 0.7\textwidth]{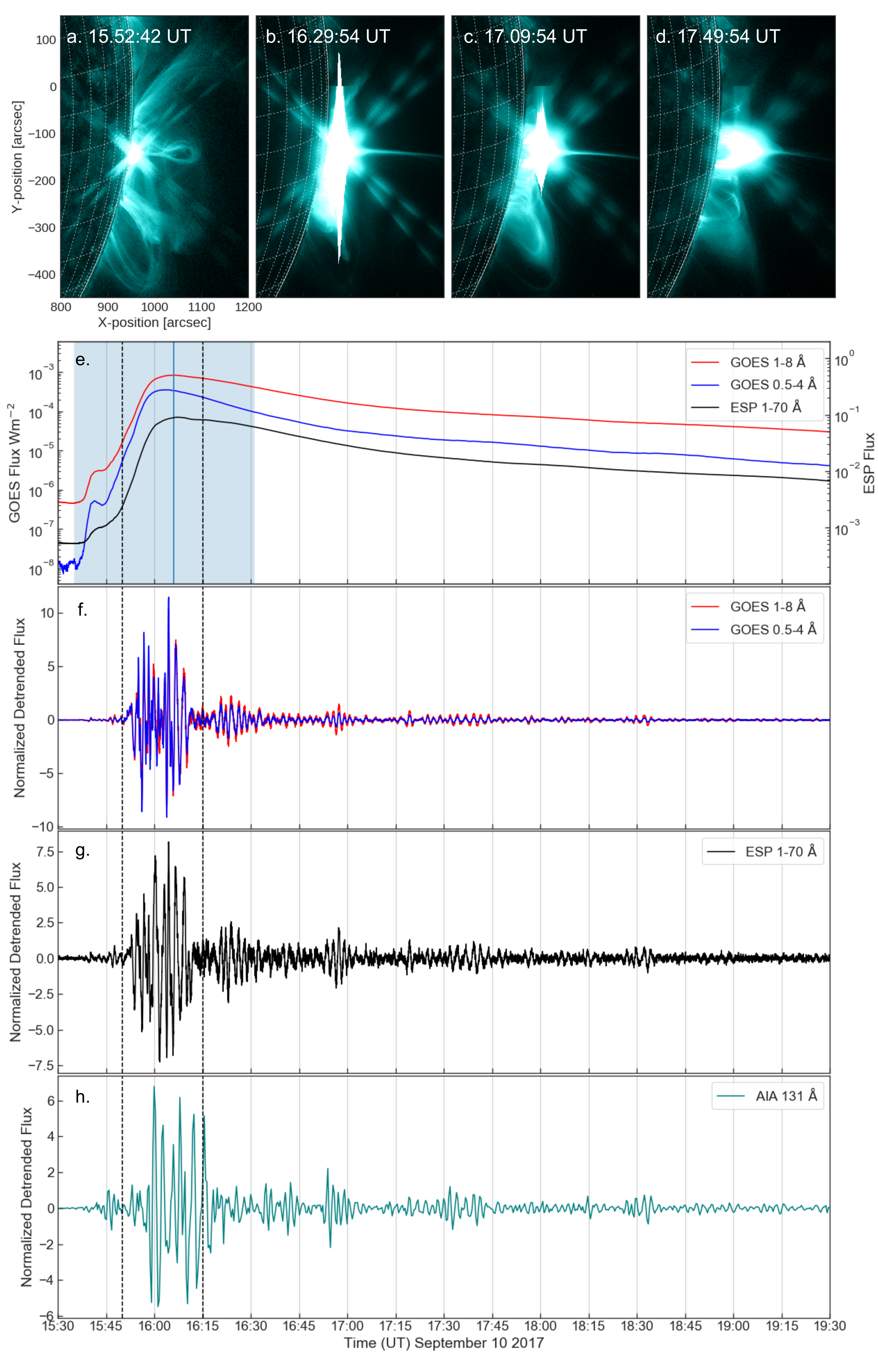}
\label{lightcurves}
\caption{The evolution of the flare seen in EUV (a-d) in the AIA 131~\AA\ passband. Detector saturation causes the blooming in (b, c) given the large magnitude of the event, and the crossed features are the diffraction effects of the detector grids. The soft X-ray lightcurves from both GOES and ESP are shown in (e). The shaded region denotes the defined start and end times and the blue vertical line is at the peak of GOES 1--8~\AA\ channel. Panels (f,g,h) show the detrended lightcurves of the two GOES channels, ESP, and AIA 131~\AA\, respectively. The detrended lightcurves are normalized by dividing by their standard deviation. Persistent QPP can clearly be identified in each channels. The vertical dashed lines in (e-h) show the impulsive phase.
}
\end{figure*}

\section{Observations} \label{sec:obs}
The flare analyzed in this paper is the X8.2 solar flare that occurred on 2017 September 10 from NOAA active region 12673. This was the second largest flare of solar cycle 24 and was accompanied by one of the fastest CMEs observed to date with a speed of $\sim$3000~km~s$^{-1}$ \citep{morosan, gopal}.  The position on the western limb of the Sun provided a unique view of the solar eruptive process, and many recent studies of different aspects of the flare have been discussed in the literature \citep[e.g.][]{omodei, gary, li_sept, warren, liu2018}.

The flare started at approximately 15:35 UT, reached a peak in the GOES 1--8~\AA\ channel at 16:06~UT and did not return to pre-flare flux levels until the next day. Figure~\ref{lightcurves}~(a--d) shows the evolution of the event as observed in the 131~\AA\ channel of AIA. The pre-eruptive plasmoid is clearly seen in (a) followed by the flare and cusp-shaped post-flaring loops (b--d). The bright linear structure extending outwards is presumably related to the reconnecting current sheet following the eruption. This feature has been identified as a current sheet in recent works \citep{warren, li_sept}, and turbulence is noted to explain why it appears wider than theoretically expected \citep{cheng2018}. Here, we will also refer to this structure as the current sheet. However, it should be noted that the structure observed in EUV may not actually be the region where reconnection itself is taking place, but more likely an emission structure perhaps associated with the reconnection outflow jets. 

In this paper we focus on the fine structure of the QPP detected in both channels (1--8~\AA\ and 0.5--4~\AA) of the \textit{Geostationary Operational Environmental Satellite's} X-ray sensor (GOES/XRS), the 1--70~\AA\ soft X-ray channel from the Extreme Ultraviolet Spectrometer (ESP) which is part of the EUV experiment \citep{woods} onboard \textit{SDO}, and the 131~\AA\ channel of \textit{SDO}/AIA. 

Both GOES and ESP provide soft X-ray high time cadence observations of 2~s and 0.25~s, respectively. In order to enhance the signal-to-noise ratio of ESP, the signal is summed to 2~s to match the GOES data.  Figure~\ref{lightcurves}~(e) shows the lightcurves of the GOES and ESP channels. In comparison to non-thermal emission in which QPP are usually clearly evident as large modulation depths, QPP in soft X-rays constitute a small fraction ($\lesssim$ 1\%) of the total emission in this energy range and are difficult to see in the raw lightcurves. In order to show the variability more clearly, we detrend the data by subtracting a background calculated using a Savitzky-Golay smoothing filter \citep{savgol, press} of degree 3 and a window size of 200~s. This window size is chosen as it was found to appropriately highlight the pulsations of interest. Similarly other window sizes can be chosen, and the QPP can also be enhanced by taking the numerical derivative (e.g. \cite{hayes2016, simoes}). Figure~\ref{lightcurves}~(f, g) show the detrended lighcurves of the soft X-ray GOES and ESP channels, respectively. Here QPP can be clearly identified during both the impulsive and decay phases. The similarity of the GOES and ESP QPP confirms that these pulsations are of solar origin and not an instrumental feature.  

It should be noted that both GOES and ESP provide Sun-as-a-star observations and so their lightcurves contain disk-integrated flux. To confirm that the observed long duration QPP signatures are associated with the emission from the flaring region of interest, we make integrated AIA 131~\AA\ lightcurves over the field of view in Figure~\ref{lightcurves}~(a--d). To avoid loss of signal from the bleeding of saturated pixels in the images, we make sure to include all the saturated pixels in this field of view (i.e. a--d). The 131~\AA\ channel of AIA has a response to the hot Fe~{\sc xxi} (10~MK) and Fe~{\sc xxii} (16~MK) lines and hence is sensitive to soft X-ray emitting plasma at a similar temperature. The detrended lightcurve of the 131~\AA\ channel is shown in Figure 1(h). QPP signatures are also clearly evident, particularly extending late into the decay phase, confirming that these long duration persistent QPP are coming from this flaring region. The AIA 131~\AA\ detrended lightcurves
also display two extra large amplitude pulsations during the impulsive phase at approximately 16:12--16:16~UT, which are not observed in GOES or ESP. These are also observed in the AIA 193~\AA\ channel lightcurves shown in the Appendix (see Figure~\ref{fig:appen_193}), confirming that they are real. It may be possible that these are associated with the lower temperature plasma that the 131~\AA\ (0.4~MK) and 193~\AA\ (1.2~MK) channels are sensitive to.

During the impulsive phase, the QPP have a more `bursty' nature and have larger amplitude than in the decay phase. The amplitude of the pulsations in all channels is small, on the order of $\sim$1\%  of the overall emission during the impulsive phase and $\sim$0.3\% during the decay phase. What is noteworthy here is that the QPP persist for up to 3 hours after the GOES peak. Furthermore, the long-duration pulsations that extend into the decay phase are observed to have a `beat' signature, with wavepackets of larger amplitude pulsations followed by times of lower amplitude pulsations.

In order to investigate the pulsations further, we break the lightcurves up into an impulsive phase from 15:50--16:15 UT (highlighted by the vertical dashed lines in \ref{lightcurves}~(e-h)) and a decay phase from 16:15--19:00 UT. We study the QPP characteristics in these time intervals and their relation to spatial flaring features observed with AIA and \textit{RHESSI}.

\section{Results}
\subsection{Periodicity}

The presence of QPP is confirmed by periodogram analysis using the Automated Flare Inference of Oscillations (AFINO) method detailed in \cite{inglis2015, inglis2016}. The Fourier power spectra of flaring time series have an intrinsic power-law shape which must be taken into account when assessing the significance of a peak in a periodogram. When a signal is detrended, spectral components are suppressed which can lead to false detections of significant periods \citep[e.g.][]{gruber, inglis2015}. The advantage of using AFINO is that the analysis is performed on the raw lightcurves with no detrending required. 

AFINO is outlined in detail in \cite{inglis2016} but is summarized here. First the raw lightcurve is normalized by the mean and a Hanning window function is applied to account for the finite duration of the time-series. The Fourier power spectrum is then calculated and a model fit and comparison test is performed. There are three models considered to represent the power spectrum - a single power-law plus a constant (model 0), a power-law-plus-constant model with a Gaussian bump (model 1) and a broken power-law model (model 2). Model 1 represents a situation with excess power at a localized frequency - i.e. the QPP model. The maximum likelihood of the three models to fit the power spectra are calculated and then tested against each other to find which model most likely represents the data. The comparison is done via the Bayesian Information Criterion (BIC), see \cite{schwartz, inglis2015, inglis2016} for a detailed explanation. A  smaller value of BIC indicates that a model is preferred to others, and hence the $\Delta$BIC between model fits represent a way to determine which model is most appropriate.  A $\Delta$BIC$_{modelA - modelB}$ \textgreater 10 suggests a model B is strongly preferred over model A \citep{kass}.

\begin{figure*}
\centering
\includegraphics[width = 0.89\textwidth]{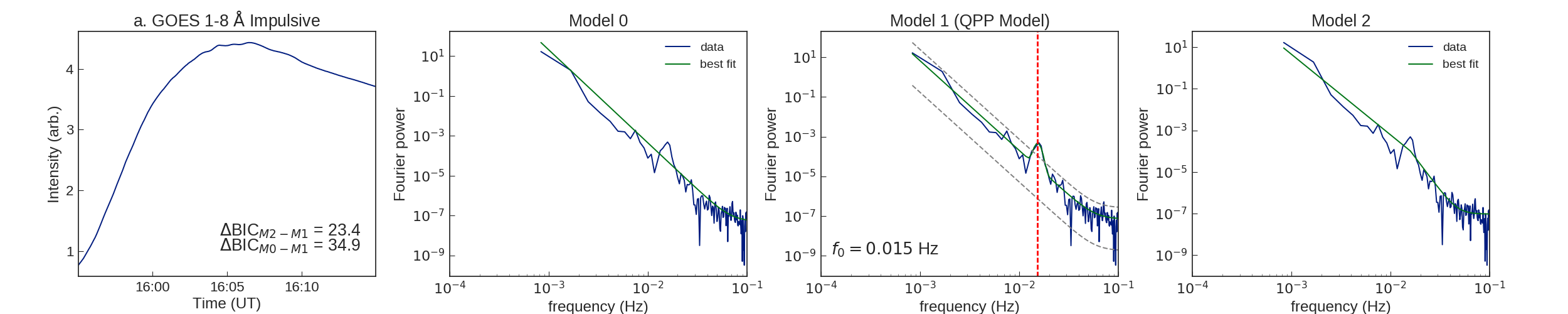}
\includegraphics[width = 0.89\textwidth]{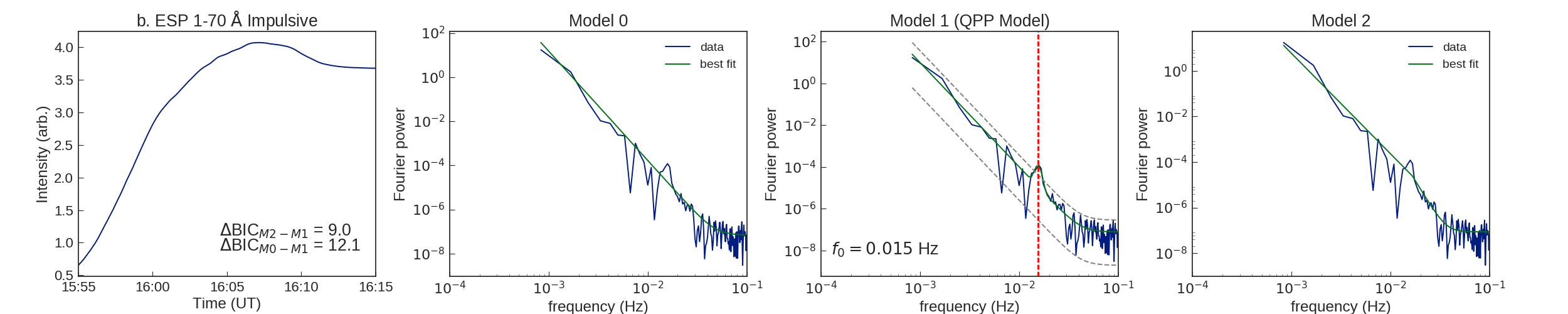}
\includegraphics[width = 0.89\textwidth]{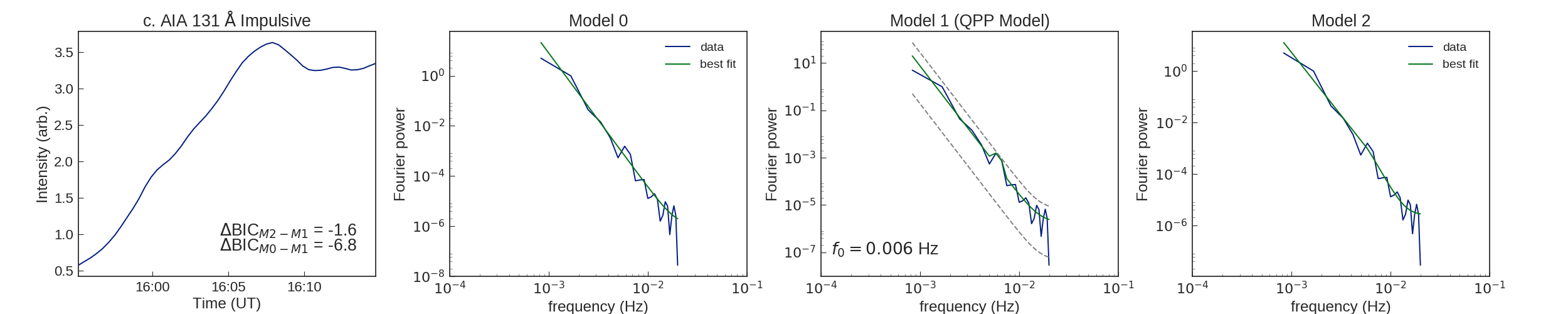}
\caption{Impulsive phase lightcurves with AFINO analysis. (a), (b) and (c) correspond to the GOES 1--8 \AA, ESP 1--70 \AA, and AIA 131 \AA\ respectively. For each, the original time-series is shown on the left panel, and the fits to each model is shown in the remaining panels. If the QPP model is preferred, a red dashed line is shown, as in (a) and (b).}
\label{afino_imp}
\end{figure*}

\begin{figure*}
\centering
\includegraphics[width = 0.89\textwidth]{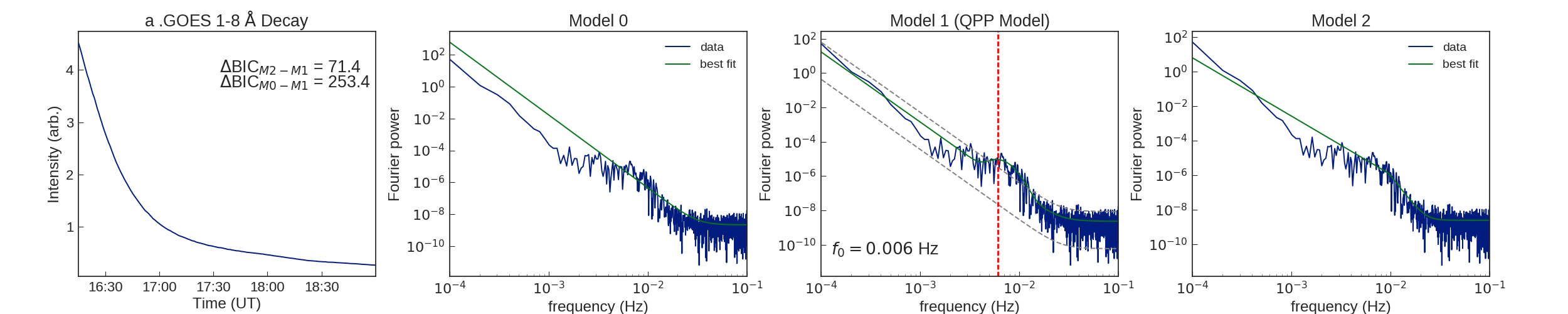}
\includegraphics[width =0.89\textwidth]{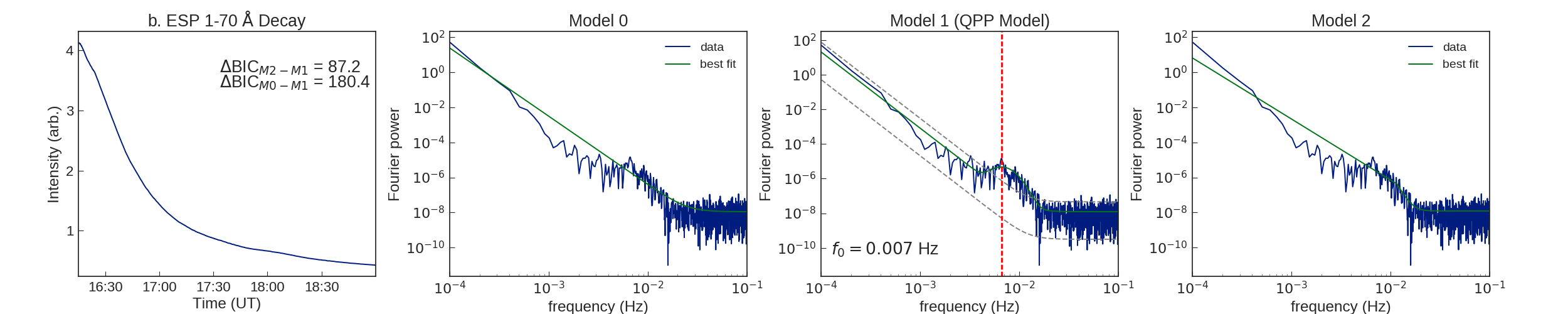}
\includegraphics[width = 0.89\textwidth]{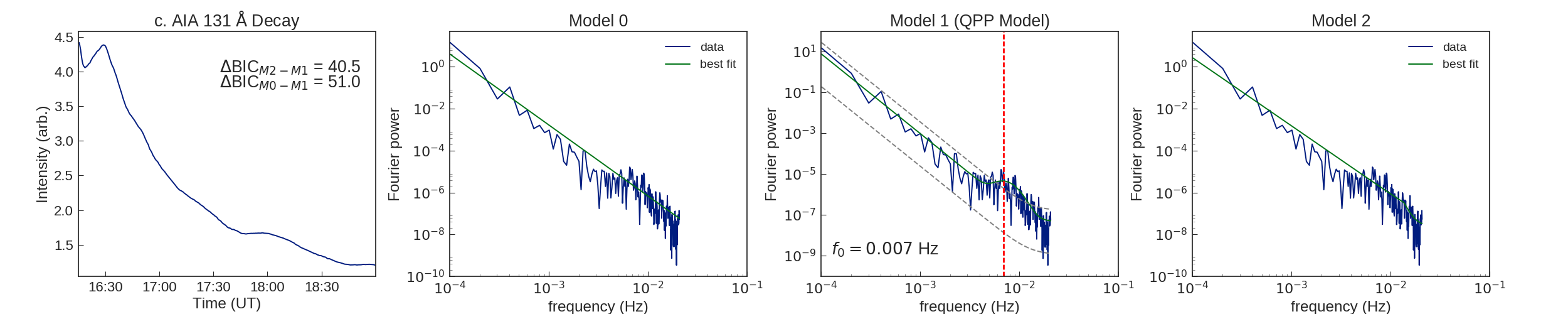}
\caption{Decay phase lightcurves with AFINO analysis. The same as Figure \ref{afino_imp}, however here all three lightcurves are best described by the QPP model with a period of 150-160s.}
\label{afino_dec}
\end{figure*}

AFINO is applied to both the impulsive and decay phases of the lightcurves separately, and the results are shown in Figure \ref{afino_imp} and \ref{afino_dec}, respectively. Each panel shows the input time-series, and the three model fits to the power spectrum. For the QPP model, the 2.5\% and  97.5\% quantiles relative to the power-law component are shown in gray. If the QPP model is preferred, a red dashed line is plotted to denote the peak of the Gaussian bump, $f_0$, which is the frequency of the detected QPP oscillation.  

It is found that during the impulsive phase, the QPP model is preferred for both the GOES channels and the ESP channel, with peak period of $\sim$65~s of the enhanced power in both these channels. The impulsive phase of the AIA 131~\AA\ lightcurve however shows no significant enhanced power in the Fourier power spectrum, and the QPP model fit to the spectrum is not found to be favored. The time cadence of the AIA 131~\AA\ lightcurve is 24~s here to ensure constant exposure time. The period of 65~s is close to the Nyquist frequency of the sampling of AIA 131~\AA\ and hence makes it difficult to detect if such a period is present. Furthermore, the BIC explicitly penalizes for a small number of data points, and given that there are much less data points for AIA compared to GOES and ESP it makes it difficult to detect a significant period at this time. This does not mean that the pulsations observed in Figure~\ref{lightcurves}~(h) are not real during the impulsive phase, just that they are not found to have a significant period with AFINO. The times between the observed peaks are estimated by eye to be $\sim$155~s. However, it again should be noted that perhaps the sampling is not permitting the detection of a shorter period. 

In the decay phase, all channels show significant enhanced power in the Fourier power spectrum, and the QPP model is found to fit best for all three instruments. The enhanced power is centered on a period of $\sim$150-160~s. This confirms the visual identification of the long duration, low amplitude QPP seen in the decay phase of Figure \ref{lightcurves}. It is notable that there is a much longer period (150~s vs 65~s) in the decay phase than in the impulsive phase observed in the soft X-ray GOES and ESP channels.  Longer timescales during the decay phase have been seen before \cite{simoes, hayes2016, dennis2017}. This may be attributed to the increase in height of emitting hot plasma in longer loops at the later stages of the flare, or perhaps a different dominant driver of the QPP signatures.

%\begin{figure*}
%\includegraphics[width = \textwidth]{pap_imp_aia.png}
%\includegraphics[width = \textwidth]{pap_dec_aia.png}
%\caption{AFINO analysis for AIA 131~\AA\ lightcurve. During the impulsive phase the QPP model fit is not preferred. However during the decay phase, the QPP model is found to best fit the power spectrum, and has the same period as both GOES and ESP.}
%\label{afino_aia}
%\end{figure*}

\begin{figure}
\centering
\includegraphics[width = 0.47\textwidth]{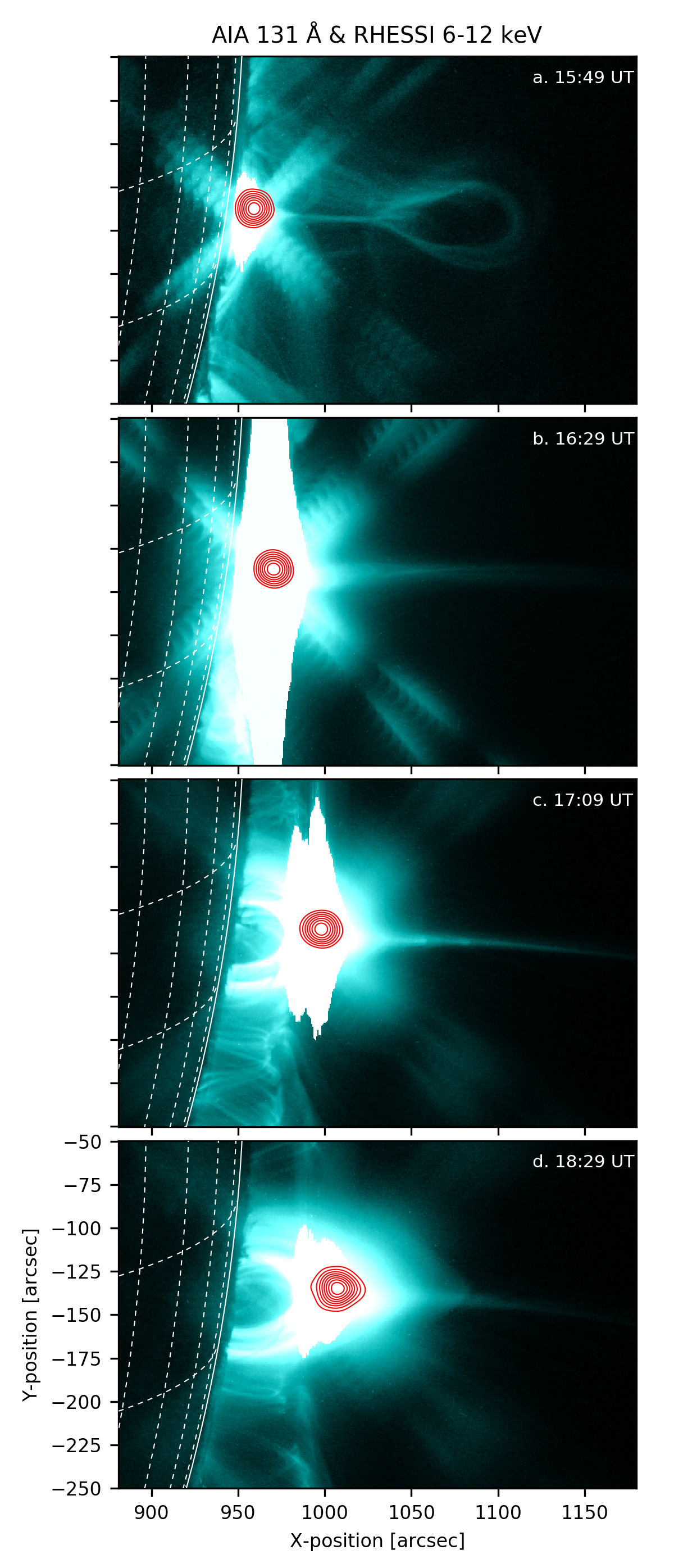}

\caption{\textit{RHESSI} contours of the 6--12~keV source overplotted on 131~\AA\ images at four stages of the event. As in Figure \ref{lightcurves}, the pre-eruption plasmoid is seen in (a). Following the eruption, the evolution of the EUV arcade is accompanied by the ascension of the soft X-ray source to higher altitudes (c--d). The 12--25~keV source is not plotted here as it shows similar evolution to the 6--12~keV, but both are shown in Figure~\ref{alt}.}
\label{rhessi_aia}
\end{figure}

\begin{figure}
\centering
\includegraphics[width = 0.47\textwidth]{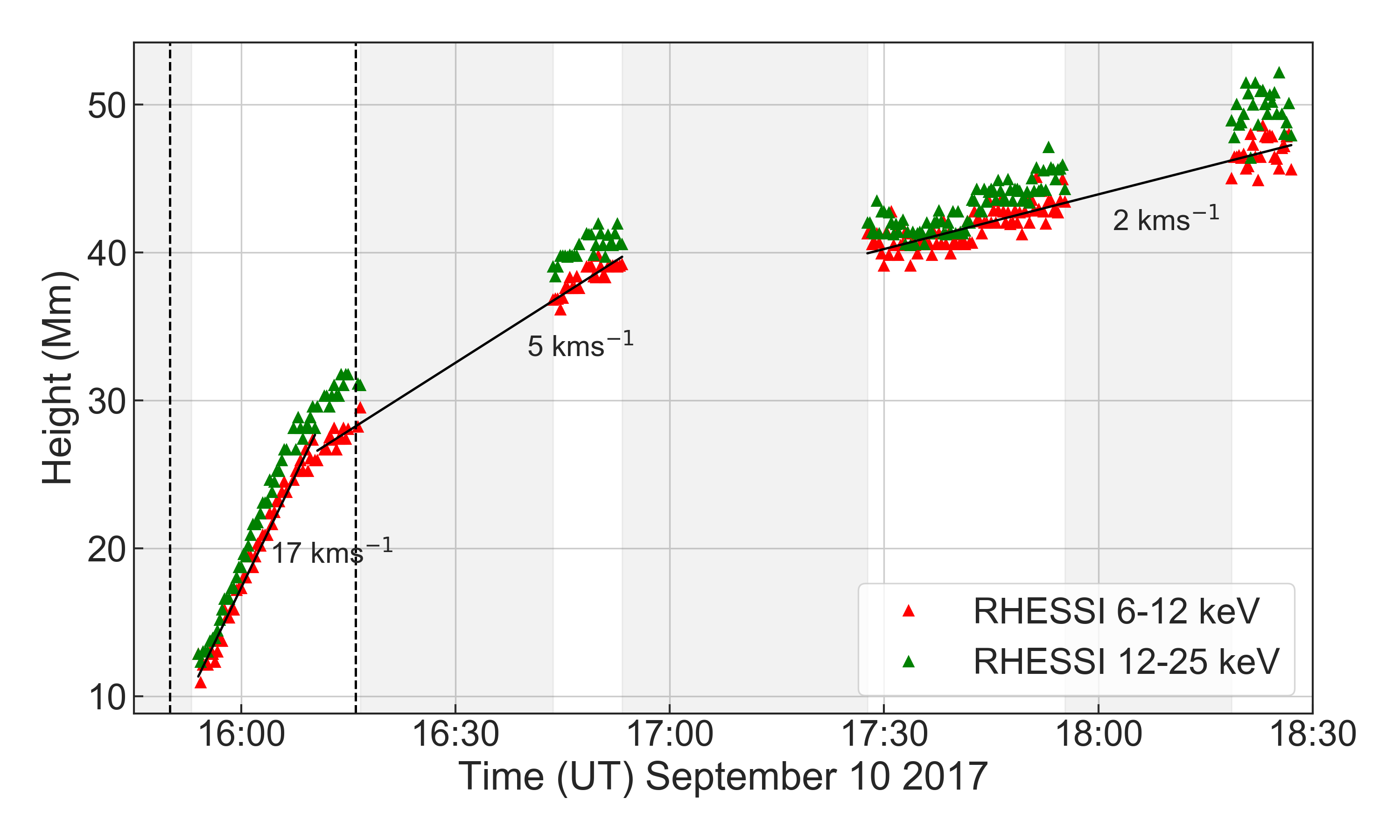}
\caption{The projected height (in the plane of sky) of the centroids of the 6--12~keV and 12--25~keV sources shown in red and green, respectively, plotted as a function of time. The velocities were determined by the fits shown by the black lines. The vertical dashed lined indicate the impulsive phase, as in Figure~\ref{lightcurves}. Gaps in the data indicate RHESSI night (shaded regions) when it is unavailable to make observations. }
\label{alt}
\end{figure}

\subsection{X-ray sources and loop length estimation}
The magnetic configuration of the solar eruptive event and its position on the limb provides an excellent view of the flare and observations of the post-flare loops face on (see Figure~\ref{lightcurves}~a--d). The flare was well observed by \textit{RHESSI}, and the on-limb location allows the soft X-ray sources to be imaged and their altitudes determined. A sequence of X-ray images in the 6--12~keV and 12--25~keV energy bands were made using the CLEAN reconstruction algorithm \citep{hurford} with detectors 3, 6 and 8, a beam-width-factor of 1, and an integration time of 20~s.

The 6--12~keV \textit{RHESSI} soft X-ray contours over-plotted on the AIA 131~\AA\ images are shown in Figure~\ref{rhessi_aia}. These soft X-ray sources are sources of hot thermal X-ray emission from the flare heated plasma. The soft X-ray sources are located at the top of the loop, presumably at a lower altitude to that of the reconnection region itself. It is observed to rise in altitude as the flare evolves, following the evolution of the EUV loops. To track the height evolution, the centroid value inside the contour of each image in the sequence is determined. The computed height of the centroids above the limb in the plane of sky is shown in Figure \ref{alt}. The gaps in the data represent periods when \textit{RHESSI} is at night and unavailable to make observations. It should be noted that the 12--25~keV source is always above the 6--12~keV source. This spatial split in energy, such that the higher energy source is situated at a higher altitude is consistent with previous works \citep{gallagher, liu}, and can be be attributed to the fact that higher loops are newly energized and thus hotter, whereas lower loops are cooling into the 6--12~keV bandpass. It should be noted that the 12--25~keV source may include some non-thermal emission together with the thermal soft X-ray emission.

The soft X-ray source rises throughout the flare, evolving rapidly during the impulsive phase with a velocity of 17~km~s$^{-1}$, slowing down to $\sim$5~km~s$^{-1}$ just after the impulsive phase, and then to 2~km~s$^{-1}$ later in the decay phase. This motion is similar to that reported by \cite{gallagher}. However, here the impulsive phase velocity of 17~km~s$^{-1}$ is faster, but the decay phase linear speed is comparable. The higher velocity during the impulsive phase is presumably related to the extremely fast ($\sim$3000~km~s$^{-1}$) associated coronal mass ejection \citep{gopal, morosan}.

It is interesting to note that the bursty, larger amplitude, QPP associated with the impulsive phase occur during a time when the soft X-ray source is rapidly rising. The decay phase pulsations, on the other hand, occur when this evolution slows down, and perhaps may explain the more stable, lower amplitude nature of the pulsations observed during this time.

\begin{figure*}
\centering
\includegraphics[width = 0.9\textwidth]{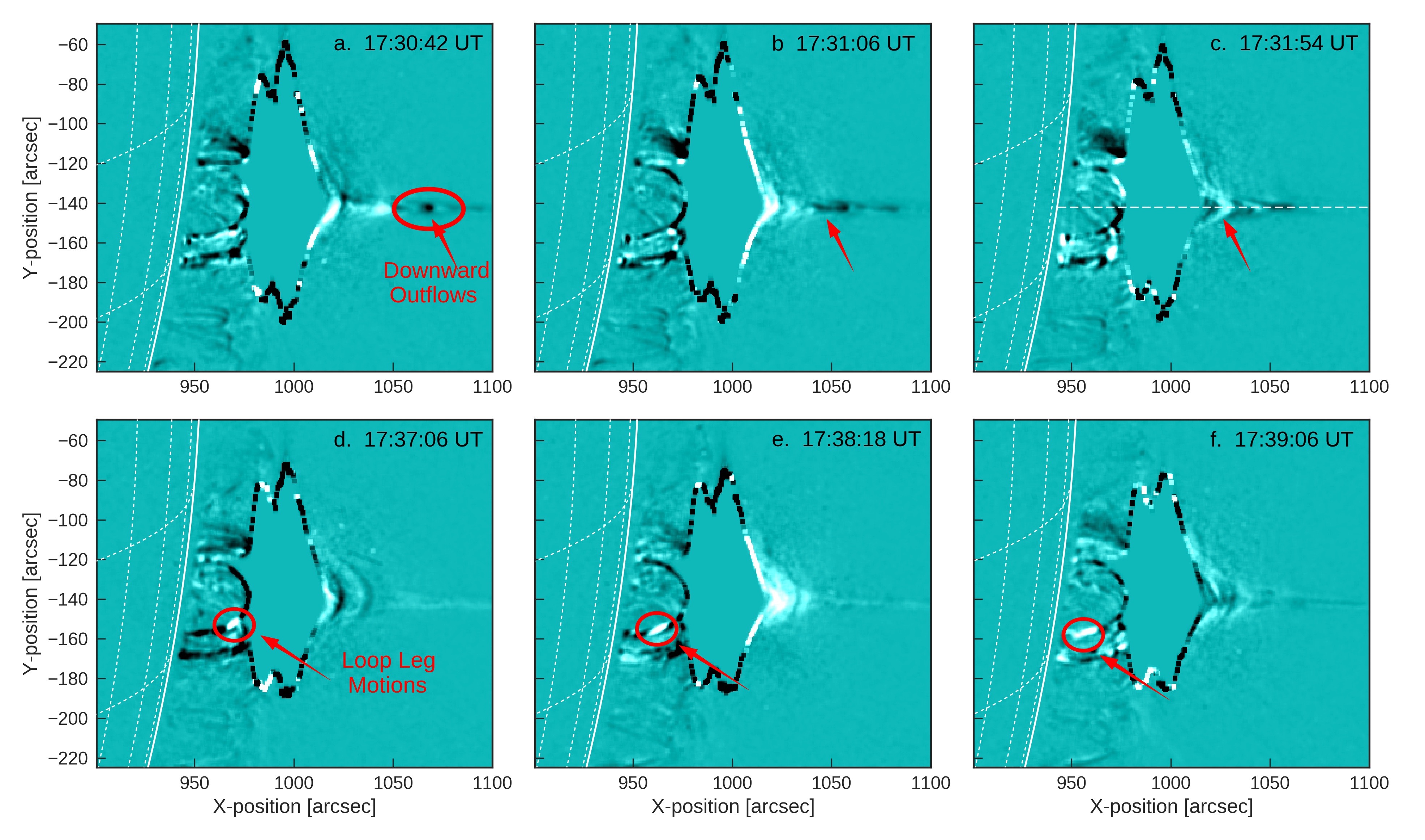}
\caption{Snapshots from animation1 of the AIA 131\AA\ running difference images. (a--c) show consecutive times to highlight the downward contracting loops/plasmoids that travel along the current sheet at impact the underlying arcade to form new post-flare loops. (d--f) show the other interesting feature noted in the movie. Loop leg motions are seen throughout the decay phase, once the saturation in the AIA pixels have subsided. }
\label{running_dif}
\end{figure*}

\subsection{AIA Loop Contractions and Downflows}
In order to enhance the features observed in the AIA 131~\AA\ images, a sequence of running difference images was produced. This allows us to investigate and track loop features and motions throughout the flare and relate them to the QPP.  The running difference movie (Animation 1) shows the plasmoid eruption with the flare emission at lower altitudes. In the early stages of the flare and during the impulsive phase, the saturated pixels dominate the running difference images and it is difficult to distinguish any features of interest. Later in the event, during the decay phase, there is clear evidence of downward retracting loops/blobs that follow a Sunward direction along the current sheet that hit the top of the newly formed cusp-shaped loops. Not only is material moving downwards towards the cusp-top of the flare, but also material is observed to be moving down the legs of the loops below the saturated pixels in the flare arcade. 

Figure~\ref{running_dif}~(a--c) and (d--f) displays three consecutive running difference images at two separate times to demonstrate motions observed in the movie and described above. In Figure~\ref{running_dif}~(a--c) we highlight the current sheet and downward directed motions presumably from the reconnection site. The red oval and arrow in (a) point to a feature that is observed to move down towards the arcade structure (b) and then form cusp-shaped loops (c). This downward moving feature is likely to be associated with supra-arcade downflows (SADs) \citep{savage} originating from newly reconnected field lines higher up along the bright linear feature that propagate towards the flare arcade apex.  In Figure~\ref{running_dif}~(d-f), the downward motions in the loop legs are shown below the saturated pixels. The red circles and arrows point to the bright feature of interest. The brightening is observed to propagate down the legs of the loop, beginning at the top (d) just below the saturated region and then travel down along the southern leg of the loop towards the southern footpoint. These motions in the loop leg are found to have a $\sim$80~s travel time from the loop top to the footpoint.

\begin{figure}
\centering
\includegraphics[width = 0.48\textwidth]{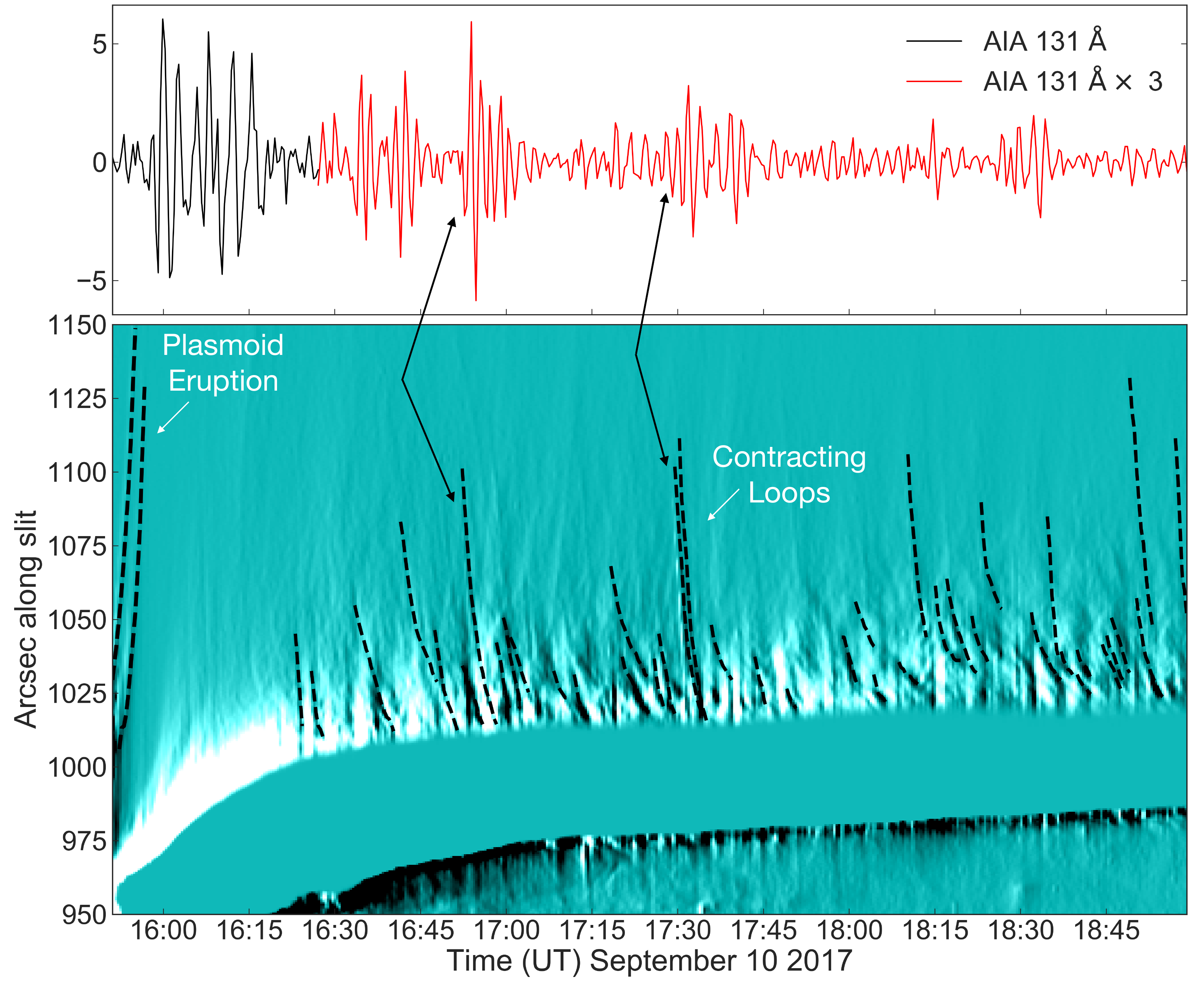}
\caption{The top panel shows the AIA 131~\AA\ detrended lightcurve. The decay phase of the lightcurve (marked in red) is multiplied by a factor of 3 to make it easier to visually identify the QPP. The bottom panel displays the space-time plot produced from the slit placed along the presumed current sheet shown in Figure~\ref{running_dif}~(c). The intensity tracks that mark motions along the slit are identified and are highlighted by the dark dashed lines. Both the plasmoid eruption at the beginning of the flare can be seen, together with the continued downward contracting loops. The arrows point to relate larger amplitude QPP to features observed to propagate to top of arcade.}
\label{slit_cut}
\end{figure}

To track the downward contracting loop motion observed in Figure~\ref{running_dif}~(a--c), a simulated slit is placed along the current sheet, noted by the white dashed line in Figure~\ref{running_dif}~(c). The slit is 10'' wide and pixels are averaged over this slit to produce the space-time plot in Figure~\ref{slit_cut}. Each moving feature is observed as a dark intensity track, which we have identified by eye and fit with the black dashed markings. At the beginning of the event, the plasmoid eruption is clearly seen. Similar to \cite{liu}, we fit each feature with a simple expression for the projected height as a function of time. The eruption is observed to have an initial velocity of 120~km~s$^{-1}$ and an acceleration of approximately 2~km~s$^{-2}$. These values are consistent with the CME velocity at this time following the eruption reported by \cite{morosan}.

The large gap between $\sim$950-1000 arcsecs in Figure~\ref{slit_cut} tracks the area of saturated pixels, and this makes it difficult to determine the loop top in AIA 131~\AA\ images. The saturated pixels, however, are clearly seen to rise in altitude during the impulsive phase followed by a more stable increase in the decay phase, similar to the \textit{RHESSI} source heights in Figure~\ref{alt}. The downward contracting loops above the cusp shaped flare loops are identified as black dashed tracks in the space-time plot (Figure~\ref{slit_cut}). The velocities of these downward contracting loops are in the range of $\sim$100-200~km~s$^{-1}$ at higher altitudes, followed by a deceleration once they arrive at the flare cusp top, where an average velocity of approximately 5-20~km~s$^{-1}$ is found. These values are consistent to those discussed in \cite{savage} and \cite{liu}.  

The detrended AIA 131~\AA\ lightcurve from Figure~\ref{lightcurves}~(h) is plotted above the space-time plot to compare the observed timings of the QPP to that of the observed EUV motions along the current sheet. The decay phase portion of the lightcurve is multiplied by a factor of 3 to aid with the visual comparison, marked in red. During the impulsive phase (from 15:50~UT to approximately 16:15~UT), large amplitude QPP are observed at the same time as the rapid rise of the emitting source as new flaring loops are formed following reconnection. After this time, the altitude increase slows down and downward contracting loops/plasmoids are observed. These are clearly also observed in the Animation1. The motion of the soft X-ray source to higher altitudes in Figure~\ref{alt} suggests that the reconnection site moves higher into the corona along the current sheet and continues to release energy. The Sunward moving blobs are presumably a result of this continued magnetic reconnection along the current sheet. These downward-moving features reach the top of the flaring arcade and form new loops. Although challenging to quantitatively determine, it is suggestive that the QPP are co-existent with the features that propagate towards the top of the loops. The timing indicates that the QPP occur co-temporally with features that are first identified at higher altitudes along the current sheet, before they reach the flare loop top. Two example of this are highlighted by the black arrows in Figure~\ref{slit_cut}.

\section{Discussion}
We have identified extensive QPP in the thermal emissions from the X8.2 flare from 2017 September 10. The pulsations persist for up to three hours after the peak of the flare, providing one of the best examples of long duration QPP observed in a flaring event. The analyzed lightcurves demonstrate two regimes of QPP signatures; bursty larger amplitude pulsations in the impulsive phase with a period of 65~s, and decay phase lower amplitude pulsations that extend late after the impulsive phase has ended, with a longer characteristic timescale of 150-160~s. During a solar flare, the impulsive and decay phases are dominated by physically different processes, and it is likely that two distinct mechanisms are underpinning the production of QPP in the different phases.

The impulsive phase of the solar flare is dominated by explosive energy release and particle acceleration due to the rapid reconfiguration of the magnetic field and the plasmoid/CME eruption. The prompt rise of the soft X-ray source during this time suggests that the reconnection site moves quickly ($\sim$17~kms$^{-1} $) to higher altitudes along the current sheet. In this way, it is difficult to interpret the impulsive phase QPP in terms of standing MHD wave modes as it is unlikely that they could be supported in the flaring loops during this complex evolution. It is more likely that the QPP at this time are related to the dominant processes of energy release and particle acceleration that have some associated characteristic timescale. 

For example, the recent work of \cite{thurgood17} has shown that reconnection at a 3D null point can proceed in a time-dependent periodic fashion and also periodically excite propagating MHD waves. This inherent property of oscillatory reconnection may play a role in producing the observed QPP. Other works have suggested that QPP are related to the formation and dynamics of plasmoids in flare current sheets \citep[e.g.][]{kliem}. Moreover, \cite{guidoni} built on the work of \cite{drake} and investigated Sunward moving plasmoids formed during reconnection in the simulation of an eruptive flare. They demonstrated that magnetic islands could trap and accelerate electrons which then interact with the flare arcade to produce emission. The discrete acceleration episodes associated with the generation and interaction of magnetic islands in a flaring current sheet could result in the observed pulsations in emission. The accelerated electrons could then precipitate to the chromosphere leading to the observed QPP in the hot soft X-ray emitting plasma \citep[similar to the Neupert effect,][]{neupert}. We did not study the hard X-ray lightcurves for this flare given the fact that they were non-uniform given the gaps in the data (due to RHESSI nighttime and long duration nature of the flare) making it difficult to perform the same robust periodicity analysis. Pulsations in hard X-ray are not directly evident by eye, which can be the case of many QPP events \citep[e.g.][]{hayes2016}. However this may be due to the fact that the flare was partially occulted \citep{gary}.

Another possible scenario is that the QPP are a manifestation of reconnection jets (the fast downflows associated with magnetic reconnection) that interact with the ambient plasma above the loop top and excite local oscillations. \cite{tak2016} performed a set of 2D MHD simulations of a solar flare and studied these oscillations excited by the reconnection outflows. They found that the above-the-loop-top region was full of shocks and oscillations and the region generated quasi-periodic propagating fast magnetoacoutsic waves (QPFs). The oscillations were controlled by multiple shocks in the region produced as a result of the collision of the reconnection outflows with the reconnected flaring loops piled up below. Their study also found quasi-periodic oscillations of the termination shock strength. It has been noted that termination shocks could be a possible site for particle acceleration \citep[e.g.][]{chen}, and hence a quasi-periodic oscillation of the termination shock at the top of the flare loop could accelerate non-thermal electrons quasi-periodically, causing QPP in the soft X-ray flux also.
More recently \cite{taka2017} performed 2D simulations of magnetic reconnection that occurs below an erupting CME, and similarly found oscillations in the above-the-loop-top region, even in the case of plasmoid-driven reconnection. In this scenario, it was found during a solar eruptive event that the impulsive phase QPP occur when the CME acceleration is peaking, which is the case for this flare. 

The model of collapsing magnetic traps within the flaring loop-top region may also be a possible mechanism to explain the X-ray QPPs \citep[see][]{jak1, jak2}. In this model, reconnected at the top of a cusp-shaped structure generates a sequence of magnetic traps which are then compressed following collision with the stronger magnetic field below. The compression results in particle acceleration,  and increases the gas and magnetic pressure inside the trap. Once this pressure increases, the traps then expand and soft X-ray emitting plasma can then fill the trap. Such a compression and expansion sets up a magnetoacoustic oscillation of the trap. The periods of the periodicity in the decay phase could be interpreted as due to changes of the length scales of the oscillating traps later in the flare evolution \citep{jak2}.

Unlike the impulsive phase, when reconnection processes and magnetic configuration evolution dominate, the decay phase is characterized by a more stable configuration of post-flare loops. In this way, it is possible the QPP at this time are a manifestation of MHD wave modes supported in the post-flare loops. The AIA running difference movie (Animation 1) shows extended downward contractions of plasmoids which reach the top of the post-flare arcade and presumably form new loops. The downward contractions could excite MHD wave oscillations in the underlying arcade, resulting in density perturbations of the post-flare loops.  The density perturbations would then be observed in the hottest thermal plasma of the newly formed loops and hence are detectable in the soft X-ray and 131~\AA\ wavebands. 

\begin{figure}
\centering
\includegraphics[width = 0.5\textwidth]{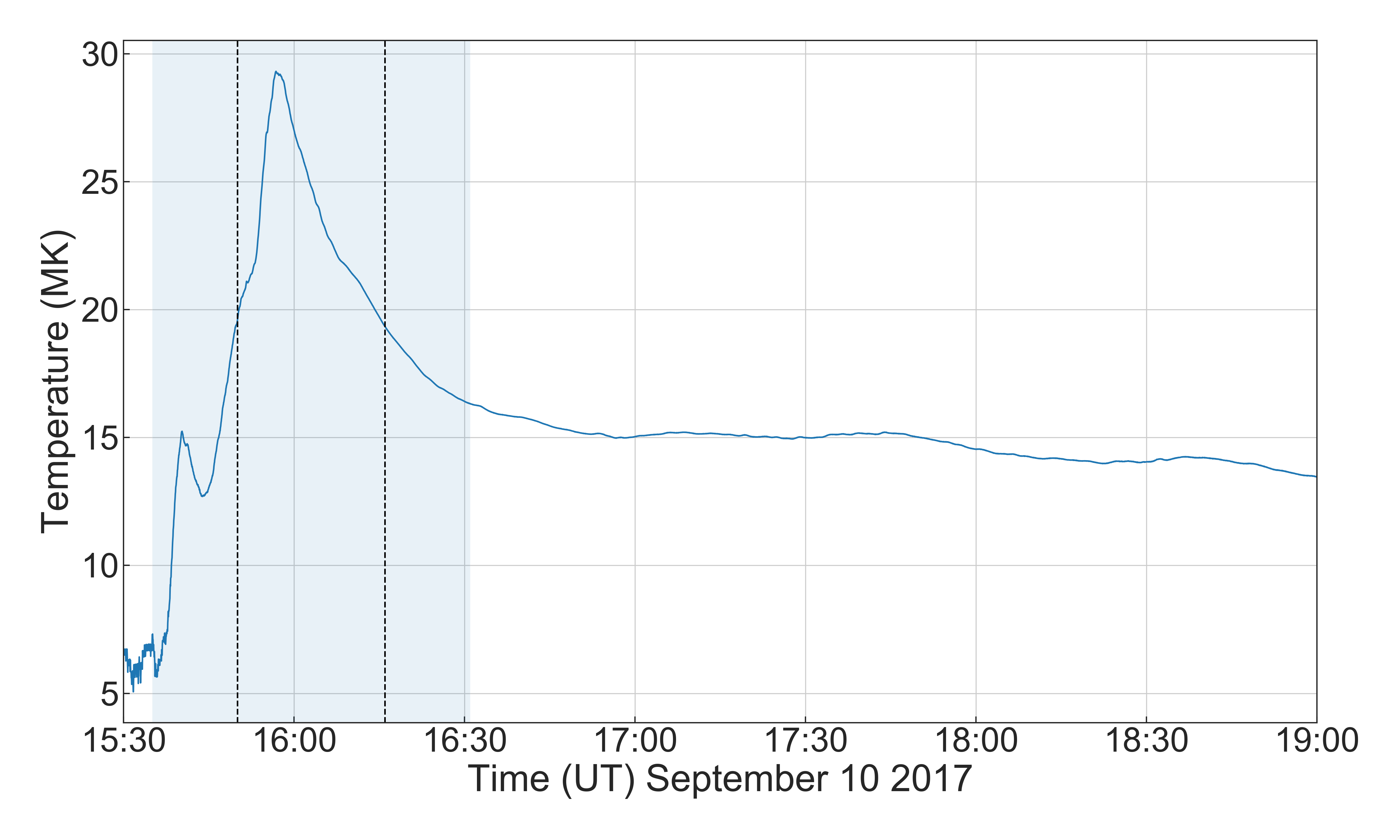}
\caption{Plot of the GOES temperature measurements throughout the flaring event. The temperature is estimated from the ratio of the short and long channels \citep{white2005}. The temperature reaches a peak of $\sim$29~K during the impulsive phase (marked by vertical dashed lines) at the maximum flux of the GOES X-ray flux. The decay phase has an approximate temperature of $\sim$15~MK, much lower than the estimated temperature from the slow magnetoacoustic interpretation (35-40~MK).}
\label{fig:TEMP_GOES}
\end{figure}

Recent numerical simulations by \cite{jelinek2017} supports this idea, as their simulation of magnetic reconnection in a current sheet found that plasmoids formed by the tearing mode instability could interact with the underlying flaring arcade and generate standing transverse MHD waves of the slow magnetoacoustic type. Slow magnetoacoustic waves can perturb the plasma density of a flaring loop and produce the observed modulation in the thermal emissions \citep[e.g.][]{wang11}. Moreover, it has been shown that an energy deposition at a flaring loop apex preferably excites the second spatial harmonic of the mode in a hot flaring loop \citep{selwa, nak04, slowmag04}, with periods in the range of 10-300~s. The period observed in this flare ($\sim$ 150-160~s during the decay phase) lies within this range and hence slow magnetoacoustic waves may be a reasonable explanation. The period, $P$ (in s), of the second harmonic of the slow magnetoacoustic wave is given by $P = 6.7L/\sqrt{T}$, where $L$ is the loop length in Mm, and $T$ is the average temperature in the loop (in MK). The loop length can be readily estimated if we assume that the loops have a semicircular shape, $L = \pi h$, where $h$ is the soft X-ray source altitude which can be determined from Figure~\ref{alt}. During the decay phase the loop lengths are in the range of 125-150~Mm. With the observed period of $150$~s, the assumption of a slow magnetoacoustic wave suggests a plasma temperature of 35-40~K. This temperature estimate is much higher than what is found from GOES temperature of $\sim$15~MK, shown in Figure~\ref{fig:TEMP_GOES}, during the decay phase. Hence this questions the validity of the slow magnetoacoustic interpretation.

Indeed, other MHD wave modes can perturb the flaring plasma. The global sausage mode is often discussed in terms of thermal QPP, given that it is a compressive wave mode and can cause a density variations within the loop to produce observed soft X-ray and EUV modulation \citep{vandoor11, tian}. However the sausage mode has a wavenumber cutoff, and can not support long period oscillations \citep{nak12}. The period here of $\sim$ 150~s is too long to be interpreted as a sausage mode oscillations of the flaring loops. 

The kink mode can also cause density perturbations when it oscillates in the vertical polarization - i.e.  up and down in the same direction of the plane of the loop rather than in the transverse horizontal motion often observed \citep{asc2011}. In the vertical polarization, the loop length can vary, moving up and down during the oscillations. This stretching and shrinking of the loop is likely to cause a density modulation of the central loop cross section, hence producing a sausage-like cross sectional and density oscillation in the loop \citep{asc2011}. For the kink-mode to move in this way, an initial excitation in the plane of the loop is required. For this event, the downward contracting loops at the top of the flare cusp could be a viable exciter. The expected changes of the soft X-ray emission in this case are expected to be small, a fraction of the overall emission, which is consistent with the observations. This interpretation has been used in the interpretation of soft X-ray pulsations for another decay phase QPP event \citep{dennis2017}. The period of the kink mode is given by $P = 2L/C_k$, where here $L$ is the loop length, and $C_k$ is the phase speed of the kink mode. The phase speed of the kink mode in coronal flaring loops can be approximated to be $\sqrt{2}  V_{A}$, where $V_{A}$ is the internal Alfv\'en speed of the loop \citep{asc_book}. Using the period values of 150~s, and a loop length of 125--150~Mm (Figure \ref{alt}), an interpretation of the QPP in terms of the kink mode oscillations suggests an Alfv\'en speed of $\sim$ 1178 - 1414~kms$^{-1}$. These are reasonable velocities for coronal loops \citep[e.g.][]{asc_book}, and perhaps suggests that the prolonged soft X-rays QPP are a manifestation of vertical kink mode oscillations in the post-flare loops. 

MHD wave modes identified in coronal loops often display distinct damped oscillatory patterns \citep{white2012}. Similarly, solar flare QPP have been reported to demonstrate damped signatures in the decay phase \citep[e.g.][]{hayes2016}. The observations of persistent QPP identified in the decay phase of this flare suggest that there is a continued renewal or excitation of MHD modes in the flaring arcade. The continued downward contractions observed in AIA 131~\AA\ along the current sheet fit this scenario and explain why continued pulsations are observed for such a long time during this flare. It should also be noted that MHD wave mode oscillations could be present during the impulsive phase. In this case however, it could be expected that the overall energy release and particle acceleration dominates and their specific oscillations cannot be identified. 

\section{Conclusions}
Long duration QPP have are investigated in the large X8.2 solar flare on 2017 September 10. Soft X-ray thermal QPP are observed throughout the impulsive phase and late into the decay phase of the flare. The decay phase pulsations are of particular interest, persisting up to 3 hours after the peak in the GOES X-ray flux. Similar to other reports of soft X-ray QPP, an evolution of the characteristic timescale of the pulsations is observed, from shorter periods in the impulsive phase ($\sim$65~s) to longer periods ($\sim$150~s) in the decay phase \citep{dennis2017, hayes2016, simoes}. These observations may reflect that this evolution is an intrinsic feature of flaring QPP, perhaps attributed to different dominant drivers of the pulsations in different stages of the flare, or instead that the timescale is related to the evolution of the loop length scales which is observed to increase throughout a flaring event.

A key finding in this work is that the decay phase pulsations co-exist with observations of extended downward motions seen in the AIA 131~\AA\ images along the current sheet that impact the top of the flaring arcade. The co-existent QPP and observed downward contractions suggest that the dynamical structure of the current sheet formation and the associated CME eruption play an important role in the generation of decay phase QPP signatures.  Furthermore, this study may help explain why no correlations were found between detected QPP periods and global active region properties such as active region size, average magnetic field strength or dipole separation in the recent statistical study performed by \cite{pugh2017}. The results here instead suggest that QPP signatures are most likely related to the loop length scales of the flare and the magnetic configuration. Future work is now required to perform a similar statistical study investigating the relationship between QPP signatures, throughout both the impulsive and decay phases, to the loop length scales and between both compact and eruptive flares.

%% If you wish to include an acknowledgments section in your paper,
%% separate it off from the body of the text using the \acknowledgments
%% command.

\acknowledgments
This work has been supported by an Enterprise Partnership Scheme studentship from the Irish Research Council (IRC) between Trinity College Dublin and ADNET Systems Inc. DEM acknowledges the Finnish Centre of Excellence in Research of Sustainable Space (Academy of Finland grant number 1312390). We would like to thank the anonymous reviewer for the helpful feedback and comments which improved this manuscript.

%% To help institutions obtain information on the effectiveness of their 
%% telescopes the AAS Journals has created a group of keywords for telescope 
%% facilities.

%% Following the acknowledgments section, use the following syntax and the
%% \facility{} or \facilities{} macros to list the keywords of facilities used 
%% in the research for the paper.  Each keyword is check against the master 
%% list during copy editing.  Individual instruments can be provided in 
%% parentheses, after the keyword, but they are not verified.

\vspace{5mm}
\facilities{GOES/XRS, RHESSI, SDO (EVE, AIA) }

%% Similar to \facility{}, there is the optional \software command to allow 
%% authors a place to specify which programs were used during the creation of 
%% the manusscript. Authors should list each code and include either a
%% citation or url to the code inside ()s when available.

\software{SunPy \cite{sunpy}}

%% Appendix material should be preceded with a single \appendix command.
%% There should be a \section command for each appendix. Mark appendix
%% subsections with the same markup you use in the main body of the paper.

%% Each Appendix (indicated with \section) will be lettered A, B, C, etc.
%% The equation counter will reset when it encounters the \appendix
%% command and will number appendix equations (A1), (A2), etc. The
%% Figure and Table counter will not reset.

\appendix

\begin{figure}
\centering
\includegraphics[width = 0.5\textwidth]{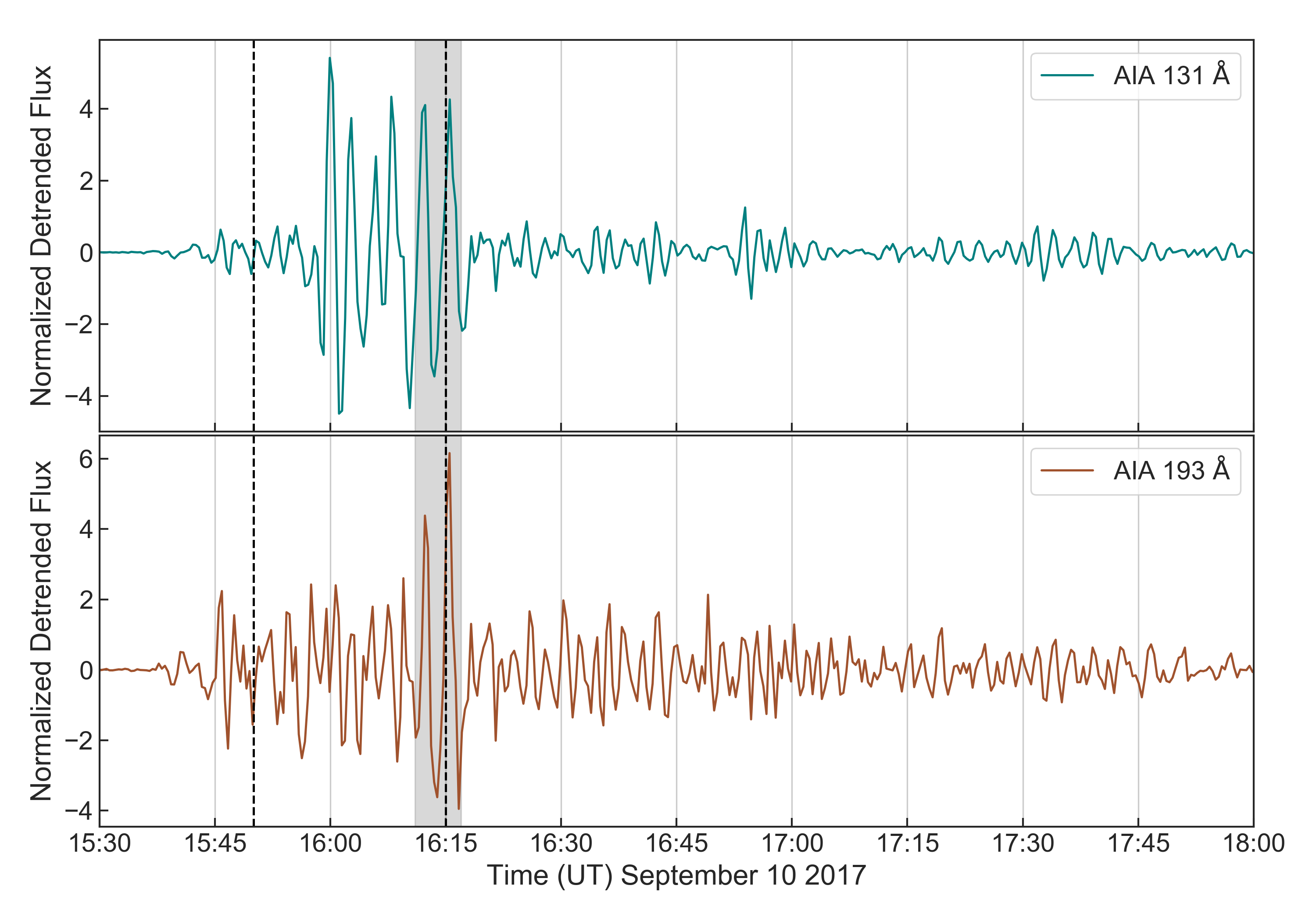}
\caption{Detrended AIA 131~\AA\ and 193~\AA\ lightcurves. The top panel shows the same as Figure~\ref{lightcurves}~(h), and the bottom panel shows the detrended lightcurve of AIA 193~\AA\ channel. The vertical dashed lines denote the impulsive phase as in Figure~\ref{lightcurves}. The grey shaded region highlight the extra two pulsations that are observed in both EUV channels but not in GOES or ESP. They are shown to demonstrate that these pulsations are real as they are observed in both channels.}

\label{fig:appen_193}
\end{figure}

\end{document}